\documentclass[aps,prb,preprint]{revtex4-2}
\usepackage{amsmath}
\usepackage{amssymb}
\usepackage{graphicx}

\begin{document}
\title{SPADExp: A photoemission angular distribution simulator directly linked to first-principles calculations}

\author{Hiroaki Tanaka}
\affiliation{Institute for Solid State Physics, The University of Tokyo, Kashiwa, Chiba 277-8581, Japan}
\email{hiroaki-tanaka@issp.u-tokyo.ac.jp}

\author{Kenta Kuroda}
\affiliation{Graduate School of Advanced Science and Engineering, Hiroshima University, Higashi-hiroshima, Hiroshima 739-8526, Japan}
\affiliation{International Institute for Sustainability with Knotted Chiral Meta Matter (WPI-SKCM${}^{2}$), Higashi-hiroshima, Hiroshima 739-8526, Japan}
\email{kuroken224@hiroshima-u.ac.jp}

\author{Tomohiro Matsushita}
\affiliation{Nara Institute of Science and Technology, Ikoma, Nara 630-0192, Japan}
\email{t-matusita@ms.naist.jp}

\begin{abstract}
We develop a software package SPADExp (simulator of photoemission angular distribution for experiments) to calculate the photoemission angular distribution (PAD), which is the momentum dependence of spectrum intensity in angle-resolved photoemission spectroscopy (ARPES).
The software can directly load the output of the first-principles software package OpenMX, so users do not need to construct tight-binding models as previous studies did for PAD calculations.
As a result, we can calculate the PADs of large systems such as quasicrystals and slab systems.
We calculate the PADs of sublattice systems (graphene and graphite) to reproduce characteristic intensity distributions, which ARPES has experimentally observed.
After that, we investigate twisted bilayer graphene, a quasicrystal showing 12-fold rotational symmetric spectra in ARPES, and the surface states of the topological insulator $\mathrm{Bi}_2\mathrm{Se}_3$.
Our calculations show good agreement with previous ARPES measurements, showing the correctness of our calculation software and further potential to investigate the photoemission spectra of novel quantum materials.
\end{abstract}


\maketitle
\section{Introduction}
Angle-resolved photoemission spectroscopy (ARPES) has been a powerful method for investigating the electronic structure of crystals \cite{RevModPhys.93.025006}.
In ARPES measurements, x-ray or ultraviolet light is irradiated on a crystal surface, and then the momenta and energies of photoelectrons are measured.
These properties are related to the band dispersion of the material, in other words, the relation between the binding energy and the Bloch wavevector.
As well as band dispersions, the momentum dependence of photoemission intensity, hereafter referred to as photoemission angular distribution (PAD), has information on the electronic structure because they are determined by the matrix element of the photoemission process related to wave functions in the solid.

Recently, the combination of first-principles calculations and ARPES measurements has revealed many emergent electronic structures.
For example, the strong topological insulator $\mathrm{Bi}_2\mathrm{Se}_3$ has cone-shaped dispersions connecting conduction and valence states on the surface \cite{Zhang2009}, which has been successfully confirmed by surface-sensitive ARPES using a vacuum ultraviolet light \cite{Xia2009}.
Light polarization dependence of ARPES spectra has revealed orbital contributions of valence bands in the iron-based superconductor FeSe \cite{PhysRevX.9.041049, Day2019}.
Furthermore, circular dichroism of PADs has visualized the pseudospin texture of topological surface states, which is also related to the spin texture \cite{PhysRevLett.107.207602}.

In this paper, we describe PAD calculations based on first-principles calculations, using newly developed software named SPADExp (simulator of photoemission angular distribution for experiments) \cite{SPADExp}.
Our software can directly load the output of the first-principles software package OpenMX \cite{PhysRevB.67.155108}, which is a significant advantage to previous studies requiring tight-binding models to discuss them \cite{PhysRevB.51.13614, DAIMON1995487, NISHIMOTO1996671, Nishimoto_1996, PhysRevB.56.7687}.
The reason for using OpenMX is that it uses localized wavefunctions as bases, which are suitable for the dipole approximation in the photoemission process \cite{PhysRevB.56.7687}.
The direct connection of our software with OpenMX enables us to calculate PADs of large systems such as quasicrystals and slab systems, as represented later; quasicrystal systems can be handled by approximating them to supercell systems.
In the case of the strong topological insulator $\mathrm{Bi}_2\mathrm{Se}_3$, we analyzed the PAD to separate the contribution of each layer on the PAD of $\mathrm{Bi}_2\mathrm{Se}_3$.
We could visualize that the topological Dirac cone state in $\mathrm{Bi}_2\mathrm{Se}_3$ is the surface state localized on the topmost layer.

The organization of the paper is as follows.
In Sec.\ \ref{Sec: Method}, we briefly explain the theory of the photoemission process.
Then the software specification is explained in Sec.\ \ref{Sec: Specification}.
Section\ \ref{Sec: Examples} is for application examples of the PAD calculations on materials and comparison with previous studies of tight-binding model calculations and ARPES measurements.
At last, we summarize our study in Sec.\ \ref{Sec: Conclusion}.
We use the Hartree atomic unit in the theoretical arguments (Sec.\ \ref{Sec: Method} and \ref{Sec: Method_details}).

\section{Overview of the calculation}
\label{Sec: Method}
The photoemission process is described by Fermi's golden rule. The excitation probability is
\begin{equation}
w_{\mathrm{FI}}=2\pi\delta(E^\mathrm{F}-E^\mathrm{I}-\omega)\Bigl|\langle \psi^\mathrm{F}|\delta H|\psi^\mathrm{I}\rangle\Bigr|^2,
\end{equation}
where $|\psi^\mathrm{I}\rangle$ and $E^\mathrm{I}$ are the wave function and the eigenenergy for the initial state, $|\psi^\mathrm{F}\rangle$ and $E^\mathrm{F}$ are those for the final state.
$\delta H$ is the time-independent part of the perturbation due to the light with the frequency $\omega$, satisfying $\delta H(t)=\delta He^{-\mathrm{i}\omega t}$.
Therefore, we need to obtain initial and final states and a perturbation term to calculate the matrix element in the excitation probability equation.
We briefly explain them in this order; see \ref{Sec: Method_details} for the details of the methodology.

An initial state is a Bloch state with a wavevector $\mathbf{k}$ and a band index $\mu$ and is represented by a linear combination of pseudo-atomic orbitals (LCAO) in OpenMX;
\begin{equation}
\psi_\mu^{(\mathbf{k})\mathrm{I}}(\mathbf{r})=\frac{1}{\sqrt{N}}\sum_{n}^N e^{\mathrm{i}\mathbf{R}_n\cdot \mathbf{k}}\sum_{i\alpha\sigma}c_{\mu,i\alpha}^{\sigma(\mathbf{k})}\phi_{i\alpha}(\mathbf{r}-\boldsymbol\tau_i-\mathbf{R}_n)|\sigma\rangle. \label{Eq: LCAO}
\end{equation}
In the equation above, $\mathbf{R}_n$ is a lattice vector, $i$ is an atom position index, $\alpha=(plm)$ is an organized orbital index with the multiplicity index $p$, angular quantum number $l$, and magnetic quantum number $m$, $\sigma$ ($\uparrow$ or $\downarrow$) is a spin index, $\phi_{i\alpha}(\mathbf{r})$ is a pseudo-atomic orbital, and $\boldsymbol\tau_i$ is an atom position vector.
In this paper, the Bloch wavevector $\mathbf{k}$ is in the extended zone scheme.
We distinguish an imaginary unit $\mathrm{i}$ and an index $i$ by capital and italic letters, respectively.
$c^{\sigma(\mathbf{k})}_{\mu,i\alpha}$ is an LCAO coefficient of the wave function and can be directly obtained from OpenMX.
Since the initial state is decomposed into wave functions attributed to atoms, we calculate the matrix element for each atom and sum it up within the independent atomic center (IAC) approximation \cite{PhysRevB.17.4573}.

We used optimized pseudo-atomic orbitals (PAOs) in first-principles calculations by OpenMX, which are generated by a linear combination of original PAOs for better convergence \cite{PhysRevB.67.155108}; the optimized basis sets are included in the OpenMX package \cite{OpenMX}.
Although original PAOs and atomic orbitals (AOs) are identical outside of a cutoff \cite{martin_2020}, AOs are appropriate for the PAD calculations because the wave functions near the nuclei are essential in the matrix element calculations.
Since all-electron calculations by ADPACK \cite{ADPACK} generates PAOs and pseudopotentials for OpenMX, they give the correspondence between original PAOs and AOs.
We calculated atomic orbital wave functions corresponding to the optimized PAOs by combining AOs with the same coefficients as the optimized PAOs.

A final state, the wavefunction of a photoelectron, is also a Bloch wave inside a crystal.
Although it is roughly a plane wave due to the high kinetic energy of a photoelectron, atomic potentials can change the eigenstate from a simple plane wave to a plane wave plus an ingoing spherical wave \cite{PhysRev.93.888}.
In addition, the wave functions of final states can rapidly decay into the bulk \cite{MOSER201729}.
The inelastic mean free paths \cite{10.1002/sia.740010103} have been employed to represent the surface sensitive property of ARPES measurements due to the decay.

The perturbation term, the electric field of the light, becomes $\mathbf{r}\cdot\mathbf{e}$ by the dipole approximation \cite{PhysRevB.56.7687}, where $\mathbf{e}$ is a unit vector representing the polarization.
The $\mathbf{r}\cdot\mathbf{e}$ term can be represented by the linear combination of $rY_{1j}(\theta,\ \varphi)$ like $\mathbf{r}\cdot\mathbf{e}=\sum_{j=-1}^1 e_j rY_{1j}(\theta,\ \varphi)$, where $e_j$ is a coefficient depending on the light polarization and direction and $Y_{1j}$ is a spherical harmonic.
This transformation enables us to separate the matrix element calculations into the spherical harmonics part and radial part.

\section{Software specification}
\label{Sec: Specification}
\begin{figure}
\centering
\includegraphics[width=120 mm]{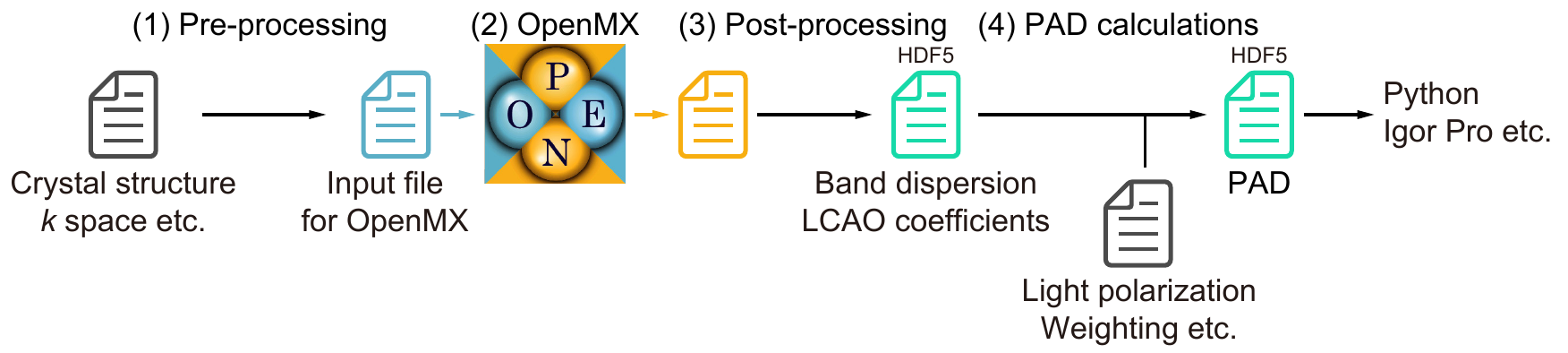}
\caption{\label{Fig: Flow} Flow of the PAD calculations by SPADExp.}
\end{figure}

We designed our software, SPADExp, as follows.
Figure \ref{Fig: Flow} summarizes the calculation flow.
First, users create a text file containing the crystal structure and the two- or one-dimensional region in the reciprocal space to calculate the band dispersion and PAD, as well as calculation parameters for OpenMX.
SPADExp can handle a spherical surface (or a curve) as well as a flat surface (or a straight line) in the reciprocal space.
The spherical surface reflects the fact that photoelectrons emitted by specified photon energy have the same $|\mathbf{k}|$ value; The $\mathbf{k}$ values of those photoelectrons are on a spherical surface.
Users may need to consider the curving effect when they use ARPES spectra with a large in-plane wavevector.
An example is represented later in the discussion of the band dispersion of graphite (Sec.\ \ref{Sec: Graphite}).
The pre-processing tool in SPADExp processes the file to generate an input file for OpenMX.
The second process is first-principles calculations by OpenMX using the created input file.
The output file of OpenMX is text-based and contains unnecessary data for the PAD calculations.
Our post-processing tool reads the output file to generate an HDF5-formatted file containing only necessary data.
The HDF5 format \cite{HDF5} is binary and can contain multiple matrices in one file.
The main routine of SPADExp loads the HDF5-formatted data and another input file specifying the light polarization, the weighting function, and so on, and then calculates the PAD.
The output of the main routine is also HDF5-formatted and can be easily visualized by graphic tools such as Python-based ones and Igor Pro.
The main routine uses the OpenMP parallelization so users can execute it efficiently on a supercomputer.

The radial wave functions of AOs and PAOs for initial states were calculated by ADPACK and are included in our software package.
Users are recommended to use AOs for initial states; we use AOs in the calculations below.

Both simple and modified plane waves are available for the final states.
Since it is difficult to calculate the potential around each atom in solids, we instead use the case of an isolated atom \cite{GOLDBERG1981285}, as previous studies \cite{DAIMON1995487,NISHIMOTO1996671,Nishimoto_1996,PhysRevB.56.7687} did.
The atomic potential and radial wave functions for each atom were obtained by the self-consistent Hartree-Fock-Slater equation \cite{herman1963atomic}; the calculation routine and the result are included in SPADExp.
Since both representations of the final states approximate the actual system, we do not have a concrete answer on which should be used.
We use modified plane waves in the calculations below other than the ones in Secs.\ \ref{Sec: Graphene} and \ref{Sec: Bi2Se3}, where we use plane waves for the comparison.
As we will show later in these subsections, the PADs can greatly change by choice of the final states when the unit cell contains multiple elements.

For the perturbation term, SPADExp adopts linear, right circular, and left circular polarizations.

The surface sensitivity of ARPES can be implemented by adding a weight function $W(L_i)$ to  the final state wave functions, $\psi_{in}^{(\mathbf{k})\mathrm{F}}$ in Eq.\ (\ref{Eq: phi_PW}) or Eq.\ (\ref{Eq: phi_Calc}), where $L_i$ is the distance of the $i$th atom from the surface.
SPADExp can use two types of weight functions.
One is the exponential decay $W(z;\ \lambda)=\exp(-z/2\lambda)$, where the square of it is related to the decay with mean free path $\lambda$.
The exponential decay is related to the case where the optical potential in many-body physics is complex and uniform \cite{HEDIN19701}.
The other is the rectangular function $W(z;\ a,\ b)$ which returns 1 if $a<z<b$ and 0 otherwise; the rectangular function can analyze the contribution of each atom layer to the photoemission intensity of bulk and surface states.

\section{Examples}
\label{Sec: Examples}
The following four subsections discuss the PADs of materials and compare them with previous ARPES measurements and model calculations.

\subsection{Graphene}
\label{Sec: Graphene}
Graphene is one of the simplest materials with sublattice, which means that the unit cell includes multiple atoms of one kind.
The band dispersion of graphene is also widely known; it has Dirac cones around the $\bar{K}$ point \cite{RevModPhys.81.109}.
Such dispersion has been investigated by measuring graphite \cite{PhysRevB.51.13614, DAIMON1995487, NISHIMOTO1996671,Nishimoto_1996} or graphene grown on a SiC surface \cite{PhysRevB.77.155303} by ARPES, based on the surface sensitivity of ARPES and the weak interlayer interaction.
Previous studies have revealed the characteristic PAD of graphene; the photoemission intensity from the lower Dirac cone is strong in the first Brillouin zone (BZ), while it is very weak out of the first BZ.
The tight-binding models have reproduced such distribution \cite{PhysRevB.51.13614, DAIMON1995487, NISHIMOTO1996671,Nishimoto_1996}, as well as the intensity calculations using the Fourier transform of the initial state wave functions \cite{PUSCHNIG2015193}.

\begin{figure}
\centering
\includegraphics[width=120 mm]{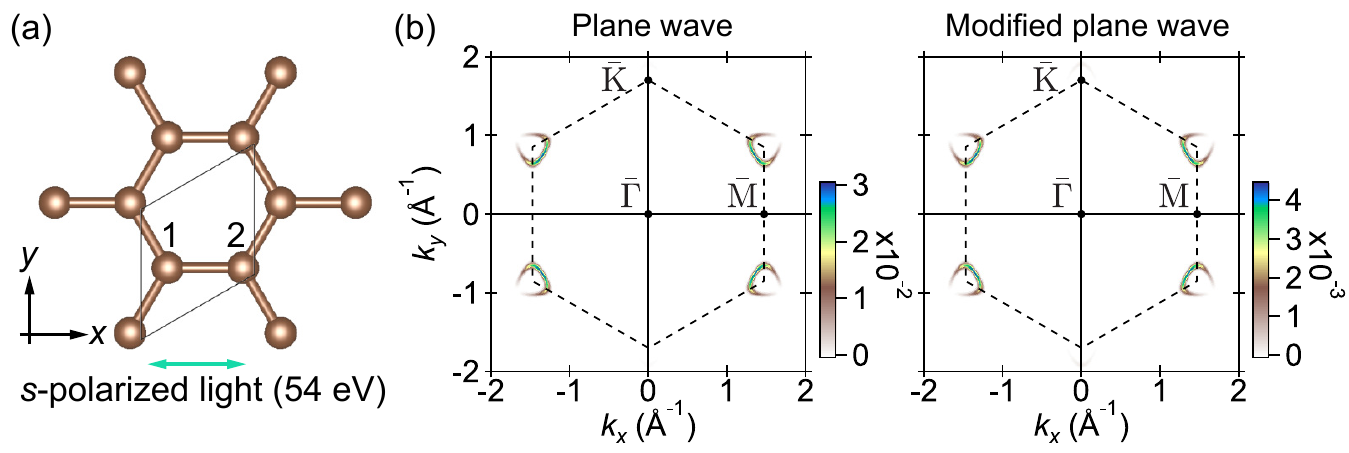}
\caption{\label{Fig: Graphene} Crystal structure and PAD of graphene. (a) Top view of the graphene unit cell. The green arrow in the bottom represents the polarization of the incident light. (b) Constant-energy map of the graphene PAD at $E_\mathrm{F}-1\ \mathrm{eV}$, calculated using plane waves [left panel] and modified plane waves [right panel] as the final states. The dashed hexagons represent the BZ of graphene.}
\end{figure}

Here, we show that SPADExp also can reproduce the PAD of graphene.
The unit cell of graphene was 20 angstroms high, with a graphene sheet placed in the center [Fig.\ \ref{Fig: Graphene}(a)].
We calculated the PAD on the plane corresponding to the photon energy of 54 eV and with the linear polarization parallel to the $\bar{\Gamma}\bar{M}$ direction.
These conditions are the same as the experimental ones \cite{NISHIMOTO1996671,Nishimoto_1996}.

The constant-energy map of the graphene PAD successfully reproduced the strong inside and weak outside intensity distributions [Fig.\ \ref{Fig: Graphene}(b)], whichever plane waves or modified plane waves were used as the final states.
In addition, the top and the bottom Dirac cones have much weaker intensity than the other four, which is also observed experimentally \cite{NISHIMOTO1996671,Nishimoto_1996}, and derived analytically \cite{DAIMON1995487}.
We find no significant change in the intensity distribution depending on the choice of the final states because the final state choice equally affects the carbon atoms in the unit cell; the change may be clearer if the unit cell contains multiple elements, as exhibited in Sec.\ \ref{Sec: Bi2Se3}.
The larger intensity in the plane wave final states than that in the modified plane wave final states is probably because the matrix element summation is performed in the complex number space due to the phase term in the modified plane wave case.

\subsection{Graphite}
\label{Sec: Graphite}
Graphite is a layered crystal of graphene sheets.
Its stacking arrangement is $AB$-type; the unit cell of graphite includes two graphene layers.
The PAD of graphite along the $k_z$ direction has been experimentally investigated by changing the photon energy.
While the unit cell with height $c$ includes two layers, the effective periodicity of the electronic structure is $c/2$ (height of one layer). Therefore ARPES measurements have observed the $k_z$ dispersion of periodicity $4\pi/c$, twice longer than the size of the reciprocal unit cell \cite{PhysRevB.97.045430}.
Such $k_z$ dispersion has also been observed in ARPES measurements of transition metal dichalcogenides \cite{PhysRevLett.119.026403, PhysRevB.105.L121102}.

\begin{figure}
\centering
\includegraphics{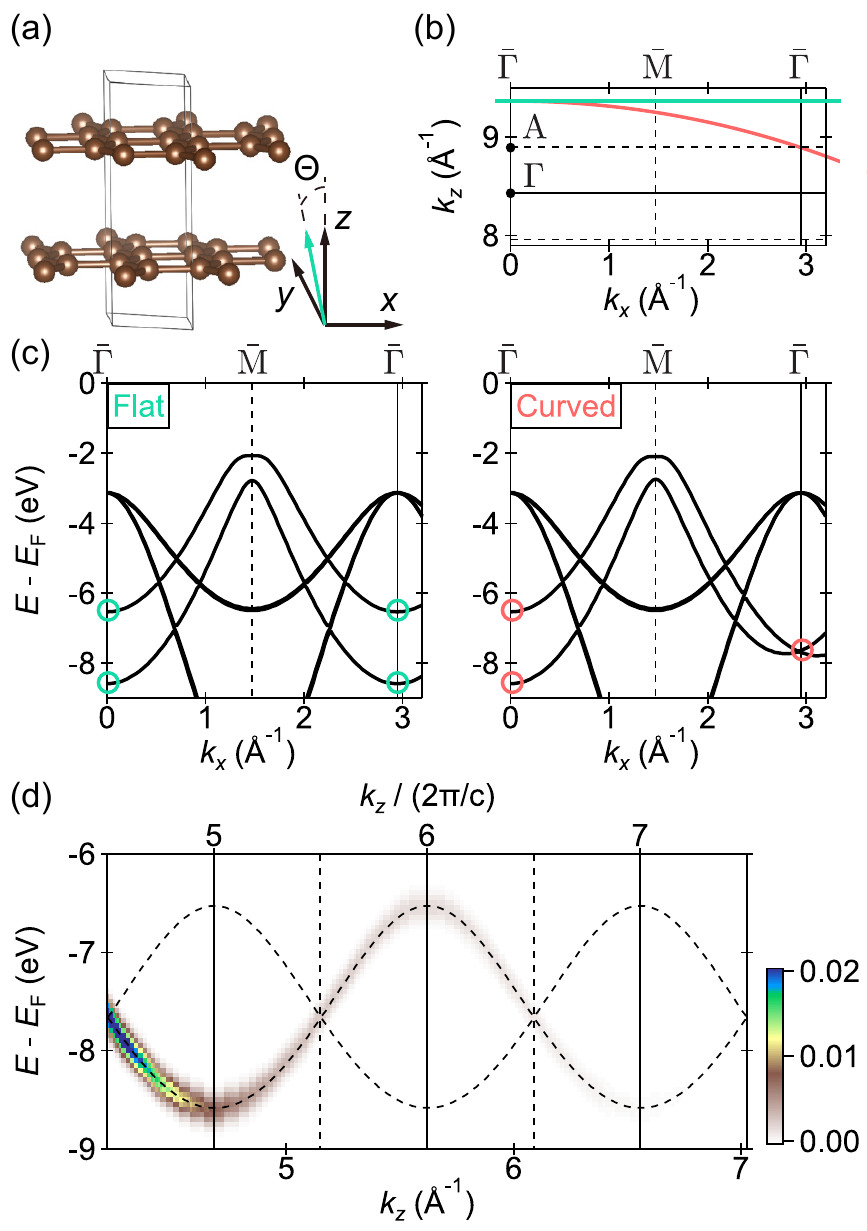}
\caption{\label{Fig: Graphite} Crystal structure, band dispersion, and PAD of graphite. (a) Unit cell of graphite. The green arrow in the right represents the polarization of the incident light, where $\Theta=30^\circ$. (b) Reciprocal lattice of graphite. The green (flat) and red (curved) curves correspond to photoemission by the light with around 300 eV photon energy. (c) Band dispersions of graphite along the green and red curves in (b). The green and red circles highlight the difference. (d) The PAD of graphite along the $k_z$ direction, the $\bar{\Gamma}$ axis at $k_x=k_y=0$.}
\end{figure}

Before discussing the PAD of graphite, we demonstrate the difference between a flat surface and a spherical surface in the reciprocal space, as presented in Sec.\ \ref{Sec: Specification}.
We calculated the dispersion of graphite $\sigma$ and $\pi$ bands along the green (flat) and red (curved) curves in Fig.\ \ref{Fig: Graphite}(b).
Both curves cross the same $\Gamma$ point at $k_x=0$, but the red curve crosses the $A$ point at the second $\bar{\Gamma}$ axis, $\pi/c$ different from the $\Gamma$ point.
Therefore, while the band dispersions at $k_x=0$ are the same between the two curves, they become different at the second $\bar{\Gamma}$ point with a nonzero $k_x$ value, as highlighted by green and red circles in Fig.\ \ref{Fig: Graphite}(c).
The reciprocal space drawn in Fig.\ \ref{Fig: Graphite}(b) corresponds to the photon energy of about 300 eV.
The result shows that we need to consider the curving effect when we see the second BZ by soft x ray ARPES.
However, the effect is negligible in ARPES measurements with the single photon energy discussed in the present paper; the photon energies are smaller than 100 eV, and they measured mainly the first BZ.
Therefore, we used flat reciprocal planes to reproduce these results.

We performed the PAD calculations for the $\pi$ bands along the $k_z$ direction, the $\bar{\Gamma}$ axis at $k_x=k_y=0$.
The $k_z$ region and the polarization [green arrow in Fig.\ \ref{Fig: Graphite}(a)] are the same as the experiment \cite{PhysRevB.97.045430}.
Figure \ref{Fig: Graphite}(d) represents the calculation result.
The PAD does not have the periodicity of $2\pi/c$, and only one of the paired bands [dashed curves in Fig.\ \ref{Fig: Graphite}(d)] has strong intensity, as experimentally observed \cite{PhysRevB.97.045430}.
The periodicity of the PAD becomes $4\pi /c$ because the effective unit cell of the electronic structure is $c/2$, the height of a graphene layer.
Within this approximation, the unit cell of the electronic structure contains one graphene layer, so the band dispersion contains only one band with a periodicity of $2\pi/(c/2)=4\pi/c$.
Furthermore, we got the same result for the question of which band has strong intensity as ARPES measurements.

\subsection{Twisted Bilayer Graphene}
\label{Sec: TBG}

Quasicrystals, which have ordered structures but do not have complete translational symmetry, have been fabricated and investigated because they can exhibit emergent symmetric structures and condensed matter properties which crystals cannot show.
Among them, twisted bilayer graphene (TBG) with a rotation angle of $30^\circ$ shows 12-fold rotational symmetric ARPES spectra \cite{doi:10.1126/science.aar8412}; such symmetry is impossible in normal crystals.
In this section, we calculate the PAD of the TBG by approximating the quasicrystal to a crystal with a much larger unit cell than that of graphene.
The direct connection of SPADExp to the first-principles software package enables us to treat such a large system without any bothering process.
Hereafter we consider the rotation angle of $90^\circ$ instead of $30^\circ$ for an easier understanding; these angles are equivalent due to the 6-fold symmetry of graphene.

\begin{figure}
\includegraphics[width=120 mm]{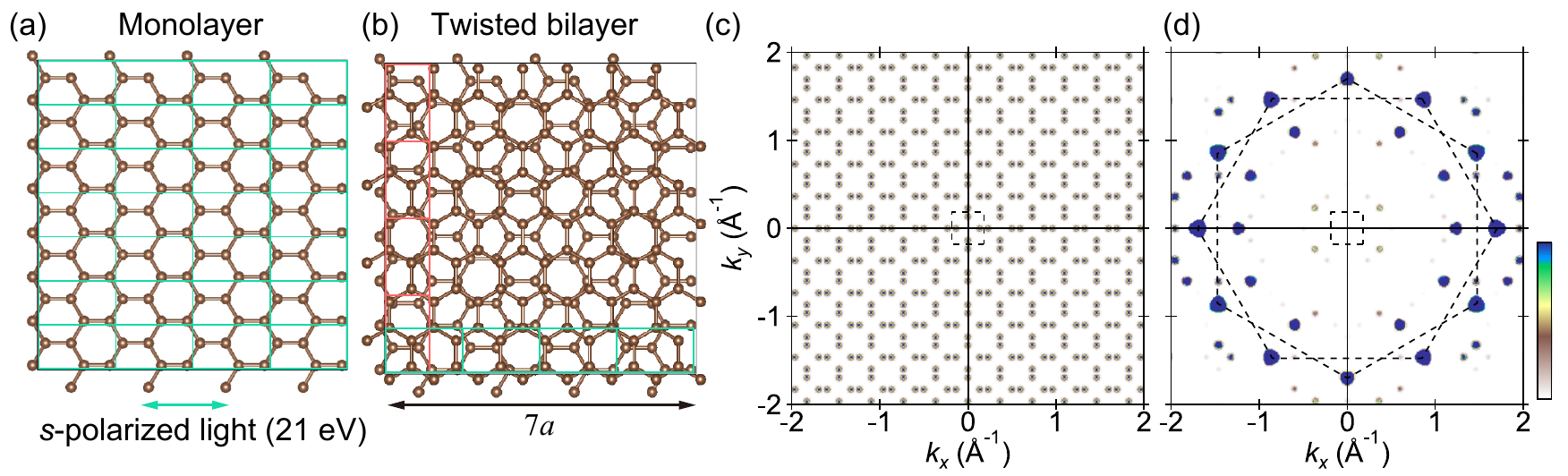}
\caption{\label{Fig: TBG} Crystal structure, band dispersion, and PAD of twisted bilayer graphene (TBG). (a) $7a\times 7a$ square filled with the rectangular unit cells of graphene. The green arrow in the bottom represents the polarization of the incident light. (b) Top view of the unit cell of the TBG. The green and red rectangles represent two ways to fill the square by graphene. (c) Constant-energy surface of the TBG band dispersion at the Fermi level. The black dashed square in the center represents the Brillouin zone (BZ) of the large unit cell in (b). (d) PAD of the TBG. The dashed hexagons represent the BZs of the graphene layers. The color scale is chosen so that the additional features in the BZ can be seen. }
\end{figure}

The origin of the quasicrystalline structure is that the rectangular unit cell of graphene [rectangles in Fig.\ \ref{Fig: TBG}(a)] is $a\times\sqrt{3}a$, and $\sqrt{3}\fallingdotseq 1.732$ is an irrational number.
Applying approximation $\sqrt{3}\simeq 7/4$, we can fill the $7a\times 7a$ square with the unit cell rectangles in two ways, as represented in Fig.\ \ref{Fig: TBG}(b); one graphene sheet is rotated by $90^\circ$ from the other.
This approximation procedure uniaxially stretches the graphene sheets by about 1 \% along the $\Gamma M$ direction.
We used a $7a\times 7a\times 20\ \text{\AA}$ unit cell with these two graphene layers in the center to calculate the TBG's electronic structure and the PAD.
Although the large unit cell has as many as 224 carbon atoms, we could perform the PAD calculations without any manual process such as the construction of the Hamiltonian matrix of the tight-binding model.

Figure \ref{Fig: TBG}(c) shows the constant-energy map of the TBG band dispersions at the Fermi energy, where the Dirac cones of the monolayer graphene touch.
The band structure is folded into the BZ of the $7a\times7a$ unit cell [black dashed square in Fig.\ \ref{Fig: TBG}(c)], so the graphene-like dispersion is lost.
Then we calculated the PAD on the plane corresponding to the excitation energy of 21 eV, as used in the previous ARPES measurements \cite{doi:10.1126/science.aar8412}; since we could not find the information on the light polarization, we used the linear polarization as represented in Fig.\ \ref{Fig: TBG}(a).
Contrary to the band dispersion, the PAD of the TBG recovers graphene-like structures, such as Dirac cones around the $\bar{K}$ points [Fig.\ \ref{Fig: TBG}(d)].
In addition to the original Dirac cones, other cones appear inside the BZ of graphene; those cones have also been observed in ARPES measurements \cite{doi:10.1126/science.aar8412}.

\subsection{Topological insulator $\mathbf{Bi}_2\mathbf{Se}_3$}
\label{Sec: Bi2Se3}
The strong topological insulator $\mathrm{Bi}_2\mathrm{Se}_3$ has an insulating electronic structure in the bulk, while the surface is metallic due to surface bands connecting the conduction and valence states \cite{RevModPhys.83.1057}.
The surface states form cone-shaped dispersion around the $\bar{\Gamma}$ point, which has been predicted by first-principles calculations \cite{Zhang2009} and observed by ARPES \cite{Xia2009}.
We investigated the PAD of $\mathrm{Bi}_2\mathrm{Se}_3$ using a slab system and the two types of weighting functions, exponential and rectangular.
The exponential weighting function is compatible with the exponential decay, which has been used in the discussion of the surface sensitivity of ARPES measurements \cite{MOSER201729}.
On the other hand, the rectangular weighting function enables us to distinguish the contribution of each layer to bulk and surface states.

\begin{figure}
\centering
\includegraphics[width=120mm]{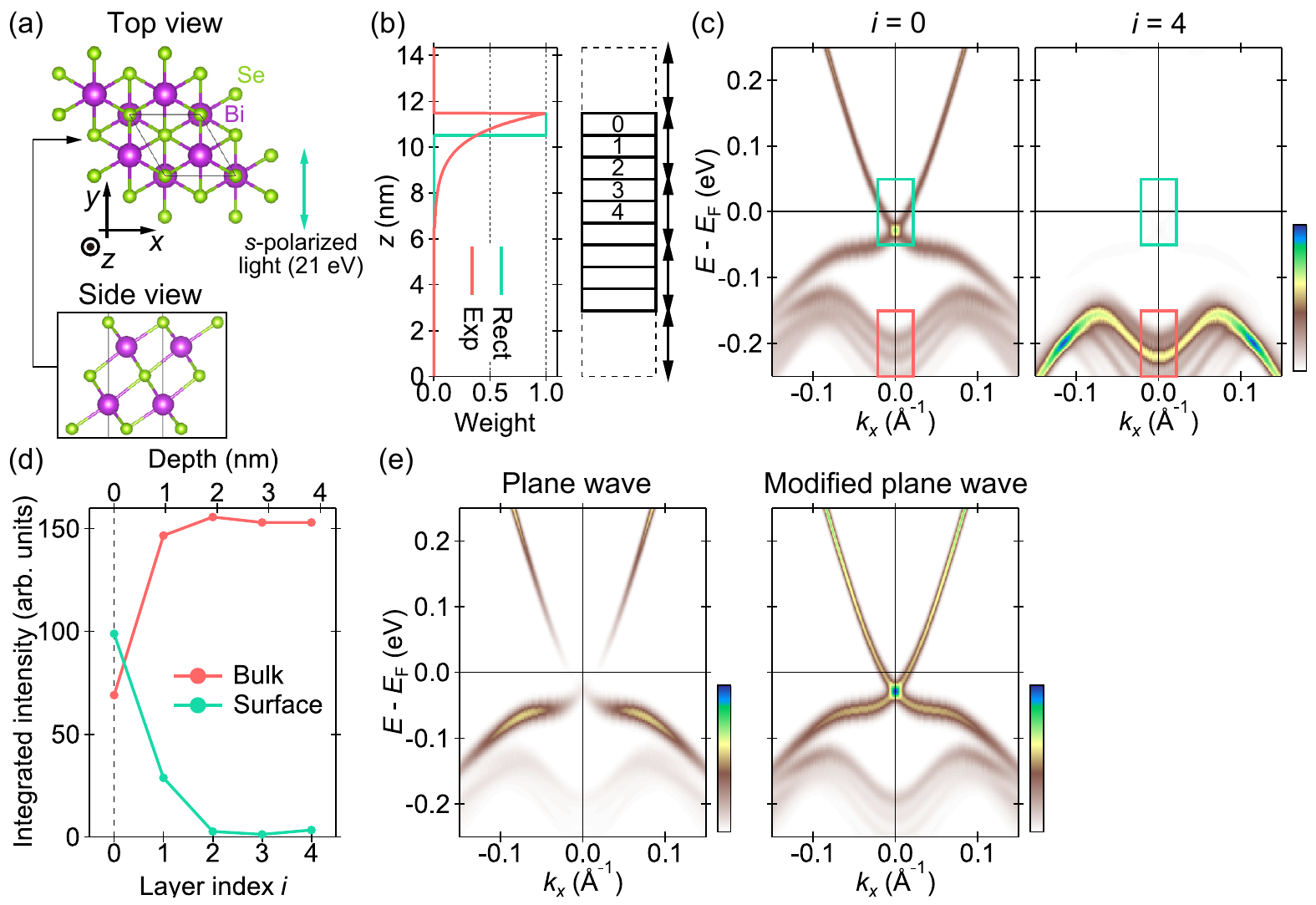}
\caption{\label{Fig: Bi2Se3} Crystal structure and PAD of $\mathrm{Bi}_2\mathrm{Se}_3$. (a) Unit cell of $\mathrm{Bi}_2\mathrm{Se}_3$. The green arrow in the right represents the polarization of the incident light. The side view shows one quintuple layer (QL). (b) Slab system composed of nine QLs. In the right figure, the black solid rectangles represent one QL, the black arrows represent a height of the unit cell containing three QLs, and the dashed rectangle represents the slab unit cell containing nine QLs and spaces. The labels 0 to 4 represent the layer index $i$. The left panel shows the weighting functions; rectangular function for the topmost QL and exponential function with $\lambda=0.5\ \mathrm{nm}$. (c) PAD of $\mathrm{Bi}_2\mathrm{Se}_3$ from the topmost layer ($i=0$) and the center layer ($i=4$). Both panels use the same color scale. (d) Layer index dependence of bulk and surface state intensities. (e) PAD of $\mathrm{Bi}_2\mathrm{Se}_3$ with the exponential weighting function, using plane waves [left panel] and modified plane waves [right panel] as the final states}.
\end{figure}

Figure \ref{Fig: Bi2Se3}(a) shows the crystal structure of $\mathrm{Bi}_2\mathrm{Se}_3$; the unit cell contains three quintuple layers (QLs).
We constructed a slab system containing nine QLs of $\mathrm{Bi}_2\mathrm{Se}_3$ [Fig.\ \ref{Fig: Bi2Se3}(b)] to discuss the surface and bulk states; We labeled the QLs from $i=0$ (topmost) to $i=4$ (center).
The rectangular weighting function [green curve in Fig.\ \ref{Fig: Bi2Se3}(b) left panel] was used to investigate the contributions of each layer to bulk and surface states.
We also used the exponential weighting function [red curve in Fig.\ \ref{Fig: Bi2Se3}(b) left panel] to compare the PAD with ARPES measurements; the mean free path $\lambda$ was set to 0.5 nm from the universal curve of the photoemission mean free path \cite{10.1002/sia.740010103}.
We employed the effective screening medium method \cite{PhysRevB.73.115407} in DFT calculations to remove dipole potential and build an isolating slab system.

The PAD from the topmost QL, calculated using the rectangular weighting function, exhibited the surface Dirac cone near the Fermi level around the $\bar{\Gamma}$ point [Fig.\ \ref{Fig: Bi2Se3}(c) left panel].
On the other hand, the PAD from the center QL showed a stronger contribution to the bulk conduction band and a much weaker contribution to the surface state [Fig.\ \ref{Fig: Bi2Se3}(c) right panel].
These results support that the cone-shaped dispersion is from surface states.
We integrated the intensities in red and green rectangles in Fig.\ \ref{Fig: Bi2Se3}(c) to analyze the contributions of each layer to bulk and surface states, respectively.
Figure \ref{Fig: Bi2Se3}(d) shows the analysis result.
The topmost QL has the strongest contribution to the surface state; the intensity rapidly decays with increasing the layer index, which shows that the electronic structure forming the Dirac cone is localized in the topmost QL.
Figure \ref{Fig: Bi2Se3}(e) shows the PADs of $\mathrm{Bi}_2\mathrm{Se}_3$ with the exponential decay, using plane waves and modified plane waves as the final states.
Since the final state modification inequally affects the bismuth and selenium atoms, the resultant PADs become more different than those of graphene, as discussed in Sec.\ \ref{Sec: Graphene}.
Since the band dispersion around the Dirac point has sufficient photoemission intensity in ARPES measurements of $\mathrm{Bi}_2\mathrm{Se}_3$ \cite{Xia2009}, the modified plane waves give closer PAD to experiments than the plane waves.
This comparison exhibits the necessity to modify the plane wave final states to reproduce PADs observed by ARPES.

\section{Conclusion}
\label{Sec: Conclusion}
We developed a software package named SPADExp to calculate PADs directly from the output of OpenMX.
Since OpenMX uses localized (pseudo-)atomic orbitals as bases, we could calculate the excitation probability with the dipole approximation.
As shown in Secs.\ \ref{Sec: Graphene}-\ref{Sec: Bi2Se3}, our PAD calculations agree well with previous ARPES studies and tight-binding calculations.
The biggest advantage of the software is that it does not need any manual process, such as constructing a tight-binding model.
Therefore, we could calculate the PADs of large systems such as the TBG quasicrystal and surface states of a slab system in Secs.\ \ref{Sec: TBG} and \ref{Sec: Bi2Se3}.
In Sec.\ \ref{Sec: Bi2Se3}, we analyzed the surface and bulk states of the topological insulator $\mathrm{Bi}_2\mathrm{Se}_3$ using the weighting functions.
We revealed that the surface states forming the Dirac cone are mostly localized in the topmost QL.
These results show the correctness of our calculations and the potential for photoemission spectrum analyses of novel quantum materials.
For example, a weak topological insulator has surface states only on the side surface \cite{Noguchi2019}, a higher-order topological insulator has edge states or corner states \cite{Noguchi2021}, and Weyl semimetals have Fermi arc states on the surface \cite{Huang2015,doi:10.1126/science.aaa9297, Lv2015,Yang2015, RevModPhys.90.015001}.
Although a system containing a large number of atoms is necessary to investigate such novel quantum materials, the combination of supercomputers and our newly developed software can handle them.

\section{Acknowledgments}
We thank Taisuke Ozaki for the discussion about OpenMX.
This work is also supported by Grants-in-Aid for JSPS Fellows (Grant No.\ JP21J20657), Grants-in-Aid for Transformative Research Areas (A) ``Hyper-Ordered Structures Science'' (Grant No.\ JP20H05884), Grants-in-Aid for Scientific Research (B) (Grants No.\ JP20H01841 and No.\ JP22H01943), and Grants-in-Aid for Scientific Research (A) (Grant No.\ JP21H04652).

\appendix
\section{Derivation of the photoemission intensity formula}
\label{Sec: Method_details}
In this section, we derive the formula to calculate the photoemission intensity following previous studies \cite{DAIMON1995487,NISHIMOTO1996671,Nishimoto_1996,PhysRevB.56.7687,PhysRevB.97.045430}.

\subsection{Dipole approximation}
In the photoemission process, the perturbation Hamiltonian due to the light irradiation $\delta H(t)$ is
\begin{align}
\delta H(t)&=\frac{1}{2}(\hat{\mathbf{p}}\cdot \mathbf{A}(t)+\mathbf{A}(t)\cdot\hat{\mathbf{p}}) \label{Eq: Perturbation-1}\\
&=\frac{A_0}{2}(\hat{\mathbf{p}}\cdot \mathbf{e}e^{\mathrm{i}\mathbf{k}^\mathrm{L}\cdot\mathbf{r}}+\mathbf{e}e^{\mathrm{i}\mathbf{k}^\mathrm{L}\cdot\mathbf{r}}\cdot\hat{\mathbf{p}})e^{-\mathrm{i}\omega t} \label{Eq: Perturbation-2}\\
&=\delta He^{-\mathrm{i}\omega t}, \label{Eq: Perturbation-3}
\end{align}
where $\hat{\mathbf{p}}$ is the momentum operator, $\mathbf{A}(t)$ is the vector potential of the light, $\mathbf{e}$ is a unit vector representing the polarization, $\mathbf{k}^\mathrm{L}$ is the wavevector of the light, and $\omega$ is the angular frequency of the light.
Since the initial state is a linear combination of localized PAOs or AOs, we can approximate $e^{\mathrm{i}\mathbf{k}^\mathrm{L}\cdot\mathbf{r}}$ by $e^{\mathrm{i}\mathbf{k}^\mathrm{L}\cdot(\boldsymbol\tau_i+\mathbf{R}_n)}$, where $\boldsymbol\tau_i+\mathbf{R}_n$ is the position of the $i$th atom.
This constant is neglected because $\mathbf{k}^\mathrm{L}$ is much smaller than $\mathbf{k}$.
By this dipole approximation and the relation $\hat{\mathbf{p}}=\frac{\mathrm{d}}{\mathrm{d}t}\hat{\mathbf{r}}$, we get
\begin{equation}
w_{\mathrm{FI}}=2\pi\delta(E^\mathrm{F}-E^\mathrm{I}-\omega)(A_0\omega)^2\Bigl|\langle\psi^\mathrm{F}|\mathbf{r}\cdot\mathbf{e}|\psi^\mathrm{I}\rangle\Bigr|^2.
\end{equation}
Previous studies have shown that the position operator $\hat{\mathbf{r}}$ should be used instead of the momentum operator $\hat{\mathbf{p}}$ for the matrix element calculations of the system with nonlocal potentials \cite{PhysRevA.3.1242,PhysRevB.44.13071}.

\subsection{Matrix element calculations}
Since the initial state is decomposed into PAOs or AOs localized at $\boldsymbol\tau_i+\mathbf{R}_n$, we calculate the matrix element for each orbital and sum it up, which uses the IAC approximation.
The sum over $\mathbf{R}_n$ gives the momentum conservation law; the matrix element can be nonzero when Bloch wavevectors of the initial and final states are different by a reciprocal lattice vector.
In the extended zone scheme, we can limit ourselves to the case where the Bloch wavevectors are identical.

At this point, we introduce atomic potentials on final states.
When we calculate the matrix element for an orbital centered at $\boldsymbol\tau_i+\mathbf{R}_n$, we set the origin of the polar coordinate system at $\boldsymbol\tau_i+\mathbf{R}_n$ for an easier understanding of equations.
As a result, the initial state wave function of the orbital specified by $\alpha=(plm)$ becomes
\begin{equation}
\phi_{\alpha in}^{(\mathbf{k})\mathrm{I}}(\mathbf{r})=e^{i\mathbf{k}\cdot\mathbf{R}_n}Y_{lm}(\theta,\ \varphi)\frac{P^{\mathrm{I}}_{ipl}(r)}{r}.
\end{equation}
The plane wave, a final state neglecting the atomic potential, is
\begin{equation}
\psi^{(\mathbf{k})\mathrm{F}}_{in}(\mathbf{r})=4\pi e^{\mathrm{i}\mathbf{k}\cdot(\boldsymbol{\tau}_i+\mathbf{R}_n)}\sum_{l^\prime m^\prime}\mathrm{i}^{l^\prime}Y_{l^\prime m^\prime}^*(\hat{\mathbf{k}})Y_{l^\prime m^\prime}(\theta,\ \varphi)j_{l^\prime}(kr). \label{Eq: phi_PW}
\end{equation}
In the equation, the partial wave expansion is applied, $\hat{\mathbf{k}}$ represents the angles $\theta$ and $\varphi$ of the vector $\mathbf{k}$, $j_{l^\prime}(x)$ is the spherical Bessel function, and $k$ is equal to $|\mathbf{k}|$.
When the potential $V_\mathrm{at}(r)$ is taken into account, the modified final state becomes the sum of the plane wave and an ingoing spherical wave like
\begin{equation}
\psi_{in}^{(\mathbf{k})\mathrm{F}}(\mathbf{r})=4\pi e^{\mathrm{i}\mathbf{k}\cdot(\boldsymbol{\tau}_i+\mathbf{R}_n)}\sum_{l^\prime m^\prime}\mathrm{i}^{l^\prime}e^{-\mathrm{i}\delta_{il^\prime}}Y_{l^\prime m^\prime}^*(\hat{\mathbf{k}})Y_{l^\prime m^\prime}(\theta,\ \varphi)\frac{P_{il^\prime}^\mathrm{F}(r)}{r}.  \label{Eq: phi_Calc}
\end{equation}
Derivation of the above form is as follows.
We can assume that $V_\mathrm{at}(r)$ is very localized around $r=0$ due to the screening effect of valence electrons \cite{Ashcroft1976}. 
Therefore, we define $r_0$ so that $V_\mathrm{at}(r)=0$ if $r>r_0$ and the radial wave function outside of $r_0$ should have the following asymptotic form;
\begin{equation}
P^\mathrm{F}_{il^\prime}(r)\rightarrow\frac{1}{k}\sin(kr-l\pi/2+\delta_{il^\prime}). \label{Eq: asymp_Pat}
\end{equation}
The asymptotic form without $\delta_{il^\prime}$ is identical to that of $rj_{l^\prime}(kr)$, the solution without $V_\mathrm{at}(r)$.
Since the asymptotic form of another solution without $V_\mathrm{at}(r)$ is obtained by changing $\sin$ to $\cos$, the asymptotic form with $V_\mathrm{at}(r)$ becomes like Eq.\ (\ref{Eq: asymp_Pat}).
The Schr\"odinger equation with $V_\mathrm{at}(r)$ gives $P^\mathrm{F}_{il^\prime}(r)$, and the phase shift $\delta_{il^\prime}$ is determined by the behavior of $P^\mathrm{F}_{il^\prime}(r)$ at large $r$ (Eq.\ \ref{Eq: asymp_Pat}). The $e^{-\mathrm{i}\delta_{il^\prime}}$ term is multiplied to remove outgoing spherical waves.
The phase term is necessary to retain the coherence of the final state, which is a key concept of the IAC approximation.

The integration of the initial state, the final state, and the perturbation term can be separated into the spherical harmonics part and the radial part.
The spherical harmonics part is the integration of $Y_{l^\prime m^\prime}^*Y_{1j}Y_{lm}$, and the integral becomes nonzero when $l^\prime=l\pm1$ and $m^\prime=m+j$ are satisfied \cite{Stohr2007}.
This integral is related to the Gaunt coefficient \cite{HMF}, and we denote it as $g(l^\prime,\ m+j;\ l,\ m)$.
The radial part is numerically calculated.

Summing up the integral over $\boldsymbol\tau_i$, $\mathbf{R}_n$, and $\alpha$, we finally obtain
\begin{align}
\langle \psi_\sigma^{(\mathbf{k})\mathrm{F}}|\mathbf{r}\cdot\mathbf{e}|\psi_{\mu}^{(\mathbf{k})\mathrm{I}}\rangle =4\pi\sqrt{N}\sum_{i\alpha}\sum_{l^\prime}^{l\pm 1}\sum_{j=-1}^1& (-\mathrm{i})^{l^\prime} Y_{l^\prime,m+j}(\hat{\mathbf{k}})e^{\mathrm{i}(\delta_{il^\prime}-\mathbf{k}\cdot\boldsymbol\tau_i)}e_j c_{\mu,i\alpha}^{\sigma(\mathbf{k})}\notag\\
&\times g(l^\prime,\ m+j;\ l,\ m)\int_0^\infty rP_{il^\prime}^{\mathrm{F}}(r)P_{ipl}^\mathrm{I}(r)\mathrm{d}r. \label{Eq: PAD}
\end{align}
Neglecting the $4\pi\sqrt{N}$ term in the beginning, we use the norm of the matrix element to draw the PAD.
Since the final states are spin-degenerated, we can calculate the matrix element for each spin separately.

\bibliography{Photoemission_angular_distribution_references}
\end{document}